\documentclass[%
reprint,
showpacs,preprintnumbers,
 amsmath,amssymb,
 aps,
]{revtex4-1}

\usepackage{graphicx}
\usepackage{dcolumn}
\usepackage{bm}
\usepackage{subfigure} 
\usepackage{color,soul}
\usepackage{ulem}

\begin{document}


\title{{Nucleation and growth of
a core-shell composite nucleus by diffusion}
}

\author{Masao Iwamatsu}
\email{iwamatsu@ph.ns.tcu.ac.jp}
\affiliation{%
Department of Physics, Faculty of Liberal Arts and Sciences, Tokyo City University, Setagaya-ku, Tokyo 158-8557, Japan
}%



\date{\today}

\begin{abstract}
The critical radius of a core-shell-type nucleus grown by diffusion in a phase-separated solution is studied.  A {\it kinetic} critical radius rather than the {\it thermodynamic} critical radius of standard classical nucleation theory can be defined from the diffusional growth equations.  It is shown that there exist two kinetic critical radii for the core-shell-type nucleus, for which both the inner core radius and the outer shell radius will be stationary.  Therefore, these two critical radii correspond to a single critical point of the nucleation path with a single energy barrier even though the nucleation looks like a two-step process.  The two radii are given by formulas similar to that of classical nucleation theory if the Ostwald-Freundlich boundary condition is imposed at the surface of the inner nucleus and that of the outer shell.  The subsequent growth of a core-shell-type post-critical nucleus follows the classical picture of Ostwald's step rule.  Our result is consistent with some of the experimental and numerical results which suggest the core-shell-type critical nucleus.
\end{abstract}

\maketitle

\section{\label{sec:sec1}Introduction}
Nucleation and growth are basic phenomena that play vital roles in the processing of various materials across many industries and in various natural phenomena~\cite{Kelton2010}.  In particular, the growth of fine particles in solution such as semiconductor quantum dots~\cite{Gorshkov2010}, bio-minerals~\cite{Meldrum2008} and other molecular crystals~\cite{Sear2012,Vekilov2012}, has attracted considerable interest recently, since such materials have many potential applications spanning the electronics to the biomedical industries.   In nucleation and growth, material transport by diffusion and its effect on the size (radius) of the critical nucleus often plays a fundamental roles in controlling the size of the products~\cite{Zener1949,Frank1950,Epstein1950,Reiss1951,Slezov2009,Wen2014}. 

In particular, growth by diffusion has been studied as it applies to various problems such as precipitation from solution~\cite{Zener1949}, liquid droplet nucleation from a supersaturated vapor~\cite{Frank1950}, vapor bubble nucleation from a supersaturated solution~\cite{Epstein1950}, and colloidal particle formation~\cite{Reiss1951} from solution. However, the nucleation and growth processes were studied separately in many cases~\cite{Robb2008,Grinin2011}.  Furthermore, nucleation by diffusion has not attracted much attention except for a recent attempt to bridge diffusion and thermodynamic evolution~\cite{Lutsko2011,Peters2011,Iwamatsu2014}.  Indeed, it is well recognized~\cite{Slezov2009} that the evolution equation known as the Zeldovich relation~\cite{Zeldovich1943, van Putten2009} from classical nucleation theory (CNT) and the diffusional growth equation are formally the same for the post-critical nucleus.  

In our previous paper~\cite{Iwamatsu2014}, we pointed out that the two evolution equations based on the thermodynamic CNT and the kinetic diffusional growth equation lead to two different definition of critical radii (i.e., {\it thermodynamic} and {\it kinetic}).  
However, except for the growth of bubbles in solution, the nucleation and growth of condensed matter proceeds via complex processes.  For example, various biomaterials such as protein crystals are not formed directly from the bulk mother solution, but rather, from within a phase-separated solution of an intermediate phase~\cite{tenWolde1997,Heijna2007,Vekilov2012,Sleutela2014} to form a core-shell composite nucleus~\cite{Iwamatsu2011,Iwamatsu2012}.  In fact, such a core-shell type nucleus has bee observed by experiments~\cite{Wang2016} and by computer simulations~\cite{tenWolde1997,Desgranges2007,Meel2008,Iwamatsu2010,Qi2015}.  Various model calculations~\cite{Kelton2000,Granasy2000,Iwamatsu2011,Santra2013,Lutsko2016} also suggest the existence of core-shell structure.  

\begin{figure}[htbp]
\begin{center}
\includegraphics[width=0.9\linewidth]{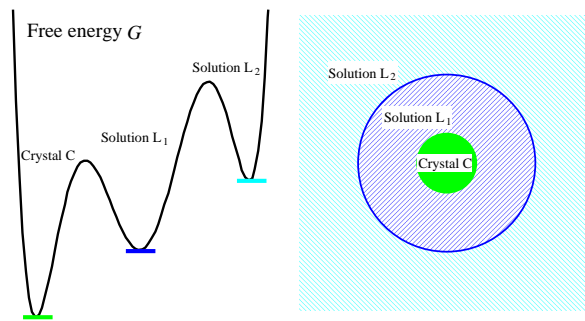}
\caption{
A model of a core-shell-type composite nucleus in a three-phase system.  The phase transition occurs from metastable solution L$_{2}$ to stable crystal C via the intermediate metastable solution L$_{1}$ (L$_{2}$$\rightarrow$L$_{1}$$\rightarrow$C) according to the Ostwald's step rule (left).  The spherical crystal nucleus C is grown from the solution L$_{1}$, which is phase-separated from the original mother solution L$_{2}$ to form a spherical shell around the crystal C.  Therefore, the critical nucleus is the composite nucleus of core-shell structure (right). }
\label{fig:1}
\end{center}
\end{figure}

In this communication, we consider the problem of nucleation and growth of the core-shell-type nucleus in solution by diffusion.  Therefore, we consider the two-step nucleation in a three-phase system~\cite{Heijna2007} according to the Ostwald's step rule~\cite{Ostwald1897,Santra2013} as shown in Fig.~\ref{fig:1}.  Initially, the pre-critical nucleus (embryo) of intermediate metastable phase (metastable dense solution L$_{1}$) nucleates within the metastable mother solution L$_{2}$.  Next, the stable crystal C nucleates from the intermediate metastable solution L$_{1}$.  However, there will be a single composite core-shell critical nucleus with a single energy barrier~\cite{tenWolde1997,Granasy2000,Iwamatsu2011,Santra2013,Qi2015}.   In such a case, the core-shell critical nucleus forms through the diffusion of materials, which must occur in two steps; bulk diffusion in the original mother solution L$_{2}$ and the diffusion in the surrounding shell of the intermediate metastable solution L$_{1}$.

This work is complementary to our previous work~\cite{Iwamatsu2012}, in which the same problem was considered from the standpoint of dynamics governed by the Fokker-Planck equation within the framework of CNT.  Here, we consider the same problem by diffusion in real space.  In section II, we show that the evolution equations for the nucleus and shell will afford two distinct critical radius for the core nucleus and the shell.  Furthermore, {\it both} the nuclear radius and the shell radius become stationary simultaneously at a single critical point, as if there were a single activation process.  This is consistent with previous thermodynamic arguments~\cite{tenWolde1997,Iwamatsu2011}, in which only a single saddle point that corresponded to the core-shell-type nucleus was identified.  In section III, we further show that by imposing the Ostwald-Freundlich boundary condition~\cite{Peters2011,Iwamatsu2014} for this core-shell system, these two critical radii are given by formulas similar to those of CNT.  Finally, in section IV, we discuss the results of our model and conclude by pointing out that the scenario given by this model would explain, qualitatively, some experimental and numerical observations of the core-shell-type nucleus.

\section{\label{sec:sec2}Nucleation and growth of a core-shell nucleus from a two-phase solution by diffusion}

We consider a spherical core-shell-type composite nucleus composed of a stable crystal phase (C) surrounded by an intermediate metastable liquid solution (L$_{1}$) nucleated in the original metastable mother liquid solution (L$_{2}$), shown in Fig.~\ref{fig:1}.  Of course, the formation of such a core-shell-type embryo (pre-critical nuclear) needs additional thermodynamic discussion~\cite{tenWolde1997,Granasy2000,Iwamatsu2011,Santra2013}.  Here, we pay most attentions to the critical and post critical nucleus after forming such a core-shell-type embryo. We consider a two-step nucleation and growth process L$_{2}$$\rightarrow$L$_{1}$$\rightarrow$C instead of one-step direct nucleation L$_{2}$$\rightarrow$C.  The crystal nucleus C with radius $R_{\rm C}$ is surrounded by an intermediate metastable solution L$_{1}$, which is separated at a radius $R_{\rm L}$ by a boundary from the original metastable solution L$_{2}$ (Fig.~\ref{fig:1}).  The two liquid phases L$_{1}$ and L$_{2}$ are assumed to be phase-separated. 

\begin{figure}[htbp]
\begin{center}
\includegraphics[width=0.9\linewidth]{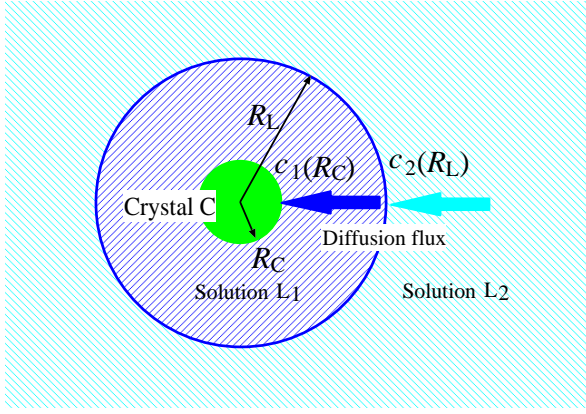}
\caption{
A diffusion flux into the core-shell composite nucleus.  A diffusion flux in the solution L$_{2}$ is absorbed into the shell of the solution L$_{1}$ and continues to diffuse in the solution L$_{1}$ into the stable crystal C.  Two fluxes in solution L$_{2}$ and L$_{1}$ must connect continuously at the boundary of radius $R_{\rm L}$. }
\label{fig:2}
\end{center}
\end{figure}

When the crystal nucleus C is grown by diffusion as shown in Fig.~\ref{fig:2}, the diffusional growth equation can be written as   
\begin{equation}
\frac{dR_{\rm C}}{dt} = -v_{\rm m} j_{R_{\rm C}},
\label{eq:D1}
\end{equation}
where $v_{\rm m}$ is the molecular volume in the nucleus C, and
\begin{equation}
j_{R_{\rm C}} = -D_{1}\left.\left(\frac{\partial c_{1}}{\partial r}\right)\right|_{r=R_{\rm C}}
\label{eq:D2}
\end{equation}
is the diffusion flux ($j_{R_{\rm C}}$) at the surface of the growing crystal nucleus, where $c_{1}$ is the concentration of solute (monomer), $D_{1}$ is the solute diffusivity in the solution L$_{1}$, and $r$ is the radial coordinate from the center of the nucleus. Note that the incoming diffusional flux must be negative ($j_{R_{\rm C}}<0$) in order to feed the nucleus to grow.   In Eq.~(\ref{eq:D1}), we have neglected the transport term due to the hydrodynamic flow~\cite{Epstein1950,Grinin2011} because we are mainly interested in the problem near the critical point where the growth of nucleus becomes stationary ($dR_{\rm C}/dt=0$).  

Now, the radius $R_{\rm C}$ of the crystal nucleus grows according to the incoming diffusion flux $j_{R_{\rm C}}$.  The concentration field $c_{i}\left(r,t\right)$, where $i=1, 2$ correspond to the solution L$_{1}$ and L$_{2}$, obeys the three-dimensional (3D) diffusion equation~\cite{Zener1949,Slezov2009,Grinin2004}
\begin{equation}
\frac{\partial c_{i}\left(r,t\right)}{\partial t}=\frac{D_{i}}{r}\frac{\partial^{2}}{\partial r^{2}}\left(r c_{i}\left(r,t\right)\right)
\label{eq:D3}
\end{equation}
for spherical symmetry, where $t$ denotes the time and $D_{i}$ denotes the solute diffusivity of the two surrounding phases, L$_{1}$ and L$_{2}$.  The steady state ($\partial c_{2}/\partial t=0$) solution for the mother solution L$_{2}$ is given by
\begin{equation}
c_{2}(r) = c_{2, \infty}-\left(c_{2, \infty}-c_{2}\left(R_{\rm L}\right)\right)\frac{R_{\rm L}}{r},\;\;\;\;R_{\rm L}<r,
\label{eq:D4}
\end{equation}
where $c_{2}\left(R_{\rm L}\right)$ is the concentration at the boundary of the two solutions L$_{1}$ and L$_{2}$  at $R_{\rm L}$ and $c_{2, \infty}=c_{2}\left(r\rightarrow \infty\right)$ is the bulk (oversaturated) concentration of the metastable mother phase L$_{2}$.  Similarly, the concentration profile of the monomers in the surrounding solution L$_{1}$ is given by
\begin{eqnarray}
c_{1}\left(r\right) &=&  \frac{R_{\rm L}R_{\rm C}\left(c_{1}\left(R_{\rm C}\right)-c_{1}\left(R_{\rm L}\right)\right)}{R_{\rm L}-R_{\rm C}}\frac{1}{r}
\nonumber \\
&&+\frac{R_{\rm L}c_{1}\left(R_{\rm L}\right)-R_{\rm C}c_{1}\left(R_{\rm C}\right)}{R_{\rm L}-R_{\rm C}},\nonumber \\
&&\;\;\;\;\;\;\;\;\;\;\;\;\;\;R_{\rm C}<r<R_{\rm L},
\label{eq:D5}
\end{eqnarray}
where $c_{1}\left(R_{\rm C}\right)$ is the concentration at the surface of the growing core nucleus at $R_{\rm C}$ and $c_{1}\left(R_{\rm L}\right)$ is the concentration at the (inner) boundary of the two solutions L$_{1}$ and L$_{2}$ at $R_{\rm L}$.  Note that these steady state concentrations do not depend on the diffusion coefficient $D_{1}$ and $D_{2}$. 

Then, the diffusion flux in the surrounding solution L$_{1}$ is given by
\begin{eqnarray}
j &=& -D_{1}\left(\frac{\partial c_{1}}{\partial r}\right) \nonumber \\
  &=& D_{1}\frac{R_{\rm L}R_{\rm C}\left(c_{1}\left(R_{\rm C}\right)-c_{1}\left(R_{\rm L}\right)\right)}{R_{\rm L}-R_{\rm C}}\frac{1}{r^{2}},\,\;\;\;\;\;R_{\rm C}\leq r\leq R_{\rm L}.
\nonumber \\
&&\label{eq:D6}
\end{eqnarray}
Similarly, the diffusion flux in the surrounding solution L$_{2}$, outside the surrounding shell of the solution L$_{1}$, is given by
\begin{eqnarray}
 j &=& -D_{2}\left(\frac{\partial c_{2}}{\partial r}\right) \nonumber \\
  &=& -D_{2}\left(c_{2, \infty}-c_{2}\left(R_{\rm L}\right)\right)\frac{R_{\rm L}}{r^{2}},\;\;\;\;R_{\rm L}\leq r.
\label{eq:D7}
\end{eqnarray}
Since the two fluxes in Eqs.~(\ref{eq:D6}) and ({\ref{eq:D7}) must be continuous at $r=R_{\rm L}$,  two alternative forms for the flux $j_{R_{\rm L}}$ at $R_{\rm L}$ are obtained:
\begin{eqnarray}
j_{R_{\rm L}}
&=& D_{1}\frac{R_{\rm C}\left(c_{1}\left(R_{\rm C}\right)-c_{1}\left(R_{\rm L}\right)\right)}{R_{\rm L}\left(R_{\rm L}-R_{\rm C}\right)} \nonumber \\
&=& -D_{2}\frac{c_{2, \infty}-c_{2}\left(R_{\rm L}\right)}{R_{\rm L}}.
\label{eq:D8}
\end{eqnarray}
Therefore, the fluxes $j_{R_{\rm C}}$, at the inner radius $r=R_{\rm C}$ is also given by two alternative forms, using either $c_{1}\left(r\right)$ or $c_{2}\left(r\right)$:
\begin{eqnarray}
j_{R_{\rm C}}
&=& D_{1}\frac{R_{\rm L}\left(c_{1}\left(R_{\rm C}\right)-c_{1}\left(R_{\rm L}\right)\right)}{R_{\rm C}\left(R_{\rm L}-R_{\rm C}\right)} \nonumber \\
&=& -D_{2}\left(\frac{R_{\rm L}}{R_{\rm C}}\right)\frac{c_{2, \infty}-c_{2}\left(R_{\rm L}\right)}{R_{\rm C}}.
\label{eq:D9}
\end{eqnarray} 
These two fluxes naturally satisfy the continuity of total flux:
\begin{equation}
4\pi R_{\rm C}^{2}j_{R_{\rm C}} = 4\pi R_{\rm L}^{2}j_{R_{\rm L}}.
\label{eq:D10}
\end{equation}
Since these fluxes $j_{R_{\rm C}}$ and $j_{R_{\rm L}}$ must be negative for the post critical nucleus, two inequalities $c_{2,\infty}>c_{2}\left(R_{\rm L}\right)$ and $c_{1}\left(R_{\rm L}\right)>c_{1}\left(R_{\rm C}\right)$ hold from Eqs.~(\ref{eq:D8}) and (\ref{eq:D9}).

Since the diffusion flux $j_{R_{\rm C}}$ at the crystal nuclear surface is incorporated into the growing nucleus, the radius $R_{\rm C}$ of the nucleus grows according to Eq.~(\ref{eq:D1}).  On the other hand the diffusion flux $j_{R_{\rm L}}$ at the surface of the L$_{1}$-L$_{2}$ boundary at $R_{\rm L}$ will be incorporated into both the surrounding solution L$_{1}$ and the nucleus C, so we must consider the conservation of material given by
\begin{equation}
\frac{dn_{1}}{dt}+\frac{dn_{\rm c}}{dt}=-4\pi R_{\rm L}^2j_{R_{\rm L}},
\label{eq:D11}
\end{equation}
where
\begin{equation}
n_{\rm c}=\frac{4\pi}{3v_{\rm m}}R_{\rm C}^{3}
\label{eq:D12}
\end{equation}
is the number of monomers in the nucleus C, and
\begin{equation}
n_{1} =\int_{R_{\rm C}}^{R_{\rm L}} 4\pi r^{2} c_{1}\left(r\right)dr
\label{eq:D13}
\end{equation}
is the number of monomers in the surrounding solution L$_{1}$.  Then, Eq.~(\ref{eq:D11}), combined with Eqs.~(\ref{eq:D12}) and (\ref{eq:D13}) leads to the evolution equation for $R_{\rm L}$
\begin{equation}
\frac{dR_{\rm L}}{dt}=-\frac{1}{c_{1}\left(R_{\rm L}\right)}j_{R_{\rm L} } +
\frac{R_{\rm C}^{2}\left(1-c_{1}\left(R_{\rm C}\right)v_{\rm m}\right)}{R_{\rm L}^{2}c_{1}\left(R_{\rm L}\right)}j_{R_{\rm C}},
\label{eq:D14}
\end{equation}
where we have replaced $dR_{\rm C}/dt$ by Eq.~(\ref{eq:D1}).  The first term on the right-hand side denotes the increase in the number of monomers in solution L$_{1}$ due to the incoming flux $j_{\rm R_{\rm L}}$, and the second term denotes the decrease in the number of monomers in L$_{1}$ which are incorporated into nucleus C.  Note that $c_{1}\left(R_{\rm C}\right)v_{\rm m}\ll 1$, as the molecular volume $1/c_{1}\left(R_{\rm C}\right)$ in solution L$_{1}$ at the surface of the nucleus at $R_{\rm C}$ is much larger than the molecular volume $v_{\rm m}$ in the nucleus C.

Using  Eq.~(\ref{eq:D10}), Eqs.~(\ref{eq:D1}) and (\ref{eq:D14}) can be written as
\begin{eqnarray}
\frac{dR_{\rm C}}{dt} 
&=& -v_{\rm m}j_{R_{\rm C}}
=-v_{\rm m}\left(\frac{R_{\rm L}^{2}}{R_{\rm C}^2}\right)j_{R_{\rm L}},
\label{eq:D15}
\\
\frac{dR_{\rm L}}{dt} 
&=& -v_{\rm m}\left(\frac{c_{1}\left(R_{\rm C}\right)}{c_{1}\left(R_{\rm L}\right)}\right)j_{R_{\rm L}}
\nonumber \\
&=& -v_{\rm m}\left(\frac{R_{\rm C}^{2}}{R_{\rm L}^2}\right)\left(\frac{c_{1}\left(R_{\rm C}\right)}{c_{1}\left(R_{\rm L}\right)}\right)j_{R_{\rm C}}.
\label{eq:D16} 
\end{eqnarray}
Therefore, the stationary conditions $dR_{\rm C}/dt=0$ and $dR_{\rm L}/dt=0$ for the critical radius $R_{\rm C}^{*}$ and $R_{\rm L}^{*}$ will be satisfied simultaneously when $j_{R_{\rm C}}=j_{R_{\rm L}}=0$, which leads to
\begin{subequations}
\begin{eqnarray}
c_{2}\left(R_{\rm L}^{*}\right)&=&c_{2, \infty},
\label{eq:D17a} \\
c_{1}\left(R_{\rm L}^{*}\right)&=&c_{1}\left(R_{\rm C}^{*}\right)=c_{1,\infty},
\label{eq:D17b}
\end{eqnarray}
\end{subequations}
where $c_{1,\infty}$ is a fictitious concentration when the crystal nucleus C is surrounded only by the solution L$_{1}$ and the solute concentration $c_{1}\left(r\right)$ is given by a formula similar to Eq.~(\ref{eq:D4}).  These simultaneous equations define the {\it kinetic critical radius} $R_{\rm C}^{*}$ and $R_{\rm L}^{*}$.  Therefore, there is a single core-shell-type critical nucleus which corresponds to a single saddle point~\cite{tenWolde1997,Iwamatsu2011}.  The critical nucleus having the radii $R_{\rm C}^{*}$ and $R_{\rm L}^{*}$ is surrounded by a wetting layer of uniform solution L$_{1}$ with the concentration $c_{1}\left(r\right)=c_{1, \infty}$ from Eq.~(\ref{eq:D5}), which is surrounded further by a uniform solutions L$_{2}$ with the concentrations $c_{2}\left(r\right)=c_{2, \infty}$ of the metastable mother solution from Eq.~(\ref{eq:D4}).  

We will not consider another solution $R_{\rm L}^{*}\rightarrow \infty$ derived from Eq.~(17a) because this solution corresponds to the direct nucleation from the metastable intermediate solution L$_{1}$ rather than the indirect nucleation from the mother solution L$_{2}$.  Note that the stationary condition $dR_{\rm C}/dt=0$ or $j_{R_{\rm C}}=0$ inevitably leads to the stationarity of another radius $dR_{\rm L}/dt=0$ or $j_{R_{\rm L}}=0$ because of the conservation of material in Eq.~(10).   When the inner radius $R_{\rm C}$ becomes stationary, the outer radius $R_{\rm L}$ must also becomes stationary. Therefore, the critical nucleus is always a composite nucleus of core-shell structure. 

\begin{figure}[htbp]
\begin{center}
\includegraphics[width=0.9\linewidth]{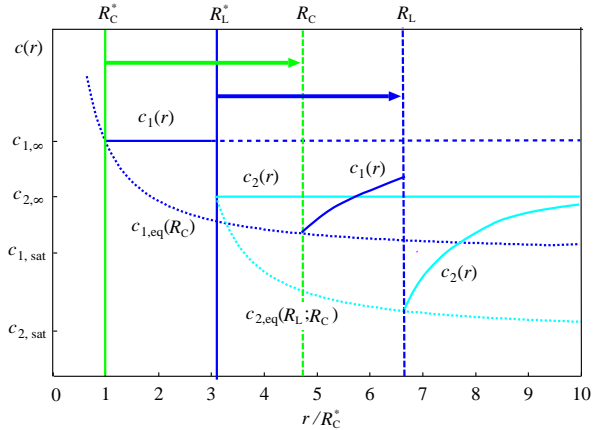}
\caption{
Concentrations $c_{1}\left(r\right)$ and $c_{2}\left(r\right)$ of monomers around the growing nucleus  as a function of distance $r$ from the center of the nucleus.  The critical nucleus having the radii $R_{\rm C}^{*}$ and $R_{\rm L}^{*}$ is surrounded by a wetting layer of uniform concentration $c_{1}\left(r\right)=c_{1}\left(R_{\rm C}^{*}\right)=c_{1}\left(R_{\rm L}^{*}\right)=c_{1, \infty}$, which is surrounded further by a uniform solutions with the concentrations $c_{2}\left(r\right)=c_{2, \infty}$ of the metastable mother solution to form a core-shell critical nucleus.  
As the post-critical nucleus begins to grow ($R_{\rm C}^{*}\rightarrow R_{\rm C}$, $R_{\rm L}^{*}\rightarrow R_{\rm L}$), the concentrations $c_{\rm 1, eq}\left(R_{\rm C}\right)$ and $c_{\rm 2, eq}\left(R_{\rm L}; R_{\rm C}\right)$ at the surface of both the core nucleus at $R_{\rm C}$ and outer boundary at $R_{\rm L}$ given by the  Ostwald-Freundlich boundary condition in Eqs.~(\ref{eq:D30}) and (\ref{eq:D32}) start to decrease.  Then, the concentrations $c_{1}\left(r\right)$ of the solution L$_{1}$ and $c_{2}\left(r\right)$ of the solution L$_{2}$ are no longer uniform and concentration gradients appears around the nucleus, according to Eqs.~(\ref{eq:D6}) and (\ref{eq:D7}), and diffusional flux ensures that the growing nucleus is fed.
}
\label{fig:3}
\end{center}
\end{figure}

In Fig.~\ref{fig:3}, we show a schematic concentration profile $c_{1}\left(r\right)=c_{1, \infty}={\rm constant}$ and $c_{2}\left(r\right)=c_{2, \infty}={\rm constant}$ for the core-shell critical nucleus with critical radii $R_{\rm C}^{*}$ and $R_{\rm L}^{*}$.  We also show a schematic concentration profile $c_{1}\left(r\right)$ and $c_{2}\left(r\right)$, which are not uniform, for a post-critical nucleus with $R_{\rm C}>R_{\rm C}^{*}$ and $R_{\rm L}>R_{\rm L}^{*}$.  The concentration profile of the post-critical nucleus will be discussed in the next section using the thermodynamic argument.

\section{\label{sec:sec3}   Ostwald-Freundlich boundary condition in a two-phase solution }

In order to find the boundary values $c_{1}\left(R_{\rm C}\right)$ and $c_{2}\left(R_{\rm L}\right)$, we will consider the chemical equilibrium of this composite core-shell nucleus.  Within the capillarity approximation, the Gibbs free energy of the composite nucleus in Fig.~\ref{fig:2} is given by 
\begin{equation}
G\left(n_{\rm c},n_{1}\right)
=-n_{\rm C}\Delta\mu_{1}-n_{1}\Delta\mu_{2}
+4\pi R_{\rm C}^{2}\gamma_{c,1}+4\pi R_{\rm L}^{2}\gamma_{1,2},
\label{eq:D18}
\end{equation}
where $\gamma_{c,1}$ and $\gamma_{1,2}$ are the surface tensions of the  C-L$_{1}$ interface at $R_{\rm C}$ and of L$_{1}$-L$_{2}$ at $R_{\rm L}$, respectively.   The number of monomers $n_{\rm C}$ in the core and $n_{1}$ in the shell are given by Eqs.~(\ref{eq:D12}) and (\ref{eq:D13}), respectively.  The chemical potential differences $\Delta \mu_{1}$ and $\Delta \mu_{2}$ of the two solutions L$_{1}$ and L$_{2}$ relative to that of the stable core nucleus C are related to the bulk solute (oversaturated) concentrations $c_{1, \infty}$ and $c_{2, \infty}$ of the two solutions L$_{1}$ and L$_{2}$ through
\begin{eqnarray}
\Delta\mu_{1} = \beta^{-1}\ln\left(c_{1, \infty}/c_{\rm 1, sat}\right),
\label{eq:D19} \\
\Delta\mu_{2} = \beta^{-1}\ln\left(c_{2, \infty}/c_{\rm 2, sat}\right),
\label{eq:D20}
\end{eqnarray}
with $\beta^{-1}=k_{\rm B}T$ and $c_{i, {\rm sat}}$ ($i=1,2$) being the saturation concentration of the two solutions which are equilibrium at the C-L$_{1}$ and L$_{1}$-L$_{2}$ flat interfaces. Note that $c_{2, \infty}=c_{2}\left(r\rightarrow\infty\right)$ (Eq.~(\ref{eq:D4})) is the concentrations of the oversaturated solution, which is larger than the saturation concentration $c_{2}\left(R_{\rm L}\rightarrow\infty\right)=c_{\rm 2, sat}$.   However, $c_{1, \infty}$ does not corresponds to $c_{1}\left(r\rightarrow\infty\right)$ (Eq.~(\ref{eq:D5})), but it is larger than $c_{1}\left(R_{\rm C}\rightarrow\infty\right)=c_{\rm 1, sat}$.

From Eq.~(\ref{eq:D18}) together with Eqs.~(\ref{eq:D12}) and (\ref{eq:D13}), the chemical potential $\mu_{\rm C}$ of the core nucleus C with radius $R_{\rm C}$ is given by
\begin{eqnarray}
\mu_{\rm C}\left(R_{\rm C}\right)&=&\frac{\partial \Delta G}{\partial n_{\rm C}}
\nonumber \\
&=&-v_{\rm m}\Delta\mu_{2}\left(c_{1}\left(R_{\rm C}^{*}\right)-c_{1}\left(R_{\rm C}\right)\right)
\nonumber \\
&&-2v_{\rm m}\gamma_{\rm c,1}\left(\frac{1}{R_{\rm C}^{*}}-\frac{1}{R_{\rm C}}\right),
\label{eq:D21}
\end{eqnarray}
where the critical radius $R_{\rm C}^{*}$ is given by an implicit equation
\begin{equation}
\Delta\mu_{1}-c_{1}\left(R_{\rm C}^{*}\right)v_{\rm m}\Delta\mu_{2}=\frac{2\gamma_{\rm c,1}v_{\rm m}}{R_{\rm C}^{*}}
\label{eq:D22}
\end{equation}
and is written formally as the classical CNT critical radius for the core nucleus
\begin{equation}
R_{\rm C}^{*}=\frac{2\gamma_{\rm c,1}v_{\rm m}}{\Delta\mu_{\rm 1,eff}},
\label{eq:D23}
\end{equation}
with
\begin{equation}
\Delta \mu_{\rm 1,eff}=\Delta\mu_{1}-c_{1, \infty}v_{\rm m}\Delta\mu_{2},
\label{eq:D24}
\end{equation}
where we have used $c_{1,\infty}=c_{1}\left(R_{\rm C}^{*}\right)$, is the effective supersaturation of the solution phase L$_{1}$ for the critical nucleus. Note that $c_{1, \infty}v_{\rm m}\ll 1$ so that $\Delta\mu_{\rm 1,eff}\simeq\Delta\mu_{1}$.  Then, Eq.~(\ref{eq:D23}) becomes exactly the CNT critical radius when the crystal C is directly nucleated from the intermediate metastable solution L$_{1}$.

Similarly, the chemical potential $\mu_{1}$ of the spherical shell of the surrounding solution L$_{1}$ with the outer radius $R_{\rm L}$ and the inner radius $R_{\rm C}$ is given by
\begin{eqnarray}
\mu_{1}\left(R_{\rm L}; R_{\rm C}\right) &=& \frac{\partial \Delta G\left(n_{\rm C},n_{1}\right)}{\partial n_{1}} \nonumber \\
&=& \frac{\Delta\mu_{2}}{c_{1}\left(R_{\rm L}\right)}\left(c_{1}\left(R_{\rm L}^{*}\right)-c_{1}\left(R_{\rm L}\right)\right) 
\nonumber \\
&&-\frac{2\gamma_{1,2}}{c_{1}\left(R_{\rm L}\right)}\left(\frac{1}{R_{\rm L}^{*}}-\frac{1}{R_{\rm L}}\right) \nonumber \\
&&+\frac{\Delta\mu_{2}}{c_{1}\left(R_{\rm C}\right)}\left(c_{1}\left(R_{\rm C}^{*}\right)-c_{1}\left(R_{\rm C}\right)\right)
\nonumber \\
&&+\frac{2\gamma_{\rm c,1}}{c_{1}\left(R_{\rm C}\right)}\left(\frac{1}{R_{\rm C}^{*}}-\frac{1}{R_{\rm C}}\right),
\label{eq:D25}
\end{eqnarray}
where, again, the critical radius $R_{\rm L}^{*}$ is given by 
\begin{equation}
R_{\rm L}^{*}=\frac{2\gamma_{1,2}}{c_{1,\infty}\Delta\mu_{2}},
\label{eq:D26}
\end{equation}
where we have used $c_{1}\left(R_{\rm L}^{*}\right)=c_{1, \infty}$ from Eq.~(17b).  The critical radius in Eq.~(\ref{eq:D26}) is given exactly by the CNT formula when a solution L$_{1}$ is nucleated from the mother solution L$_{2}$.  Note that the chemical potential $\mu_{1}$ in Eq.~(\ref{eq:D25}) depends indirectly on the core radius $R_{\rm C}$.  Then Eq.~(\ref{eq:D21}) is written as
\begin{eqnarray}
\mu_{\rm C}\left(R_{\rm C}\right)
&=&-\frac{2v_{\rm m}\gamma_{1,2}}{R_{\rm L}^{*}c_{1, \infty}}\left(c_{1, \infty}-c_{1}\left(R_{\rm C}\right)\right)
\nonumber \\
&&-2v_{\rm m}\gamma_{\rm c,1}\left(\frac{1}{R_{\rm C}^{*}}-\frac{1}{R_{\rm C}}\right),
\label{eq:D27}
\end{eqnarray}
and Eq.~(\ref{eq:D25}) is written as
\begin{eqnarray}
\mu_{1}\left(R_{\rm L}; R_{\rm C}\right) 
&=& \frac{2\gamma_{1,2}}{R_{\rm L}^{*}}\left(\frac{1}{c_{1}\left(R_{\rm L}\right)}-\frac{1}{c_{1, \infty}}\right) 
\nonumber \\
&&-\frac{2\gamma_{1,2}}{c_{1}\left(R_{\rm L}\right)}\left(\frac{1}{R_{\rm L}^{*}}-\frac{1}{R_{\rm L}}\right) \nonumber \\
&&+\frac{2\gamma_{1,2}}{R_{\rm L}^{*}}
\left(\frac{1}{c_{1}\left(R_{\rm C}\right)}-\frac{1}{c_{1, \infty}}\right)
\nonumber \\
&&+\frac{2\gamma_{\rm c,1}}{c_{1}\left(R_{\rm C}\right)}\left(\frac{1}{R_{\rm C}^{*}}-\frac{1}{R_{\rm C}}\right),
\label{eq:D28}
\end{eqnarray}
where we have used Eq.~(17b).  Equations (\ref{eq:D27}) and (\ref{eq:D28}) are the thermodynamic driving force of the growing nucleus. 

Therefore, if the surrounding solution L$_{1}$ is always in chemical equilibrium with the core nucleus at the C-L$_{1}$ interface at $R_{\rm C}$, the concentration at the interface $c_{1}\left(R_{\rm C}\right)$ is given by the equilibrium concentration $c_{\rm 1, eq}\left(R_{\rm C}\right)$, which is related to the chemical potential $\mu_{\rm C}$ of the core nucleus C given by Eq.~(\ref{eq:D27}) through
\begin{equation}
\mu_{\rm C}\left(R_{\rm C}\right)=\beta^{-1}\ln\left(c_{\rm 1, eq}\left(R_{\rm C}\right)/c_{1, \infty}\right),
\label{eq:D29}
\end{equation}
which leads to the Ostwald-Freundlich (OF) or Gibbs-Thomson equation given by
\begin{equation}
c_{\rm 1, eq}\left(R_{\rm C}\right)=c_{1, \infty}\exp\left(\beta\mu_{\rm C}\left(R_{\rm C}\right)\right).
\label{eq:D30}
\end{equation}
Then, $c_{\rm 1, eq}\left(R_{\rm C}^{*}\right)=c_{1, \infty}$ at the critical point of nucleation from Eq.~(17b) because $\mu_{\rm C}\left(R_{\rm C}^{*}\right)=0$.  The concentration $c_{\rm 1, eq}\left(R_{\rm C}\right)$ is expected to decreases monotonically as a function of $R_{\rm C}$ from Eq.~(\ref{eq:D27}).

Similarly, if the solution L$_{2}$ is in chemical equilibrium with a droplet shell of the solution L$_{1}$ with the outer radius $R_{\rm L}$ and inner radius $R_{\rm C}$, the concentration at the interface $c_{2}\left(R_{\rm L}\right)$ is given by the equilibrium concentration $c_{\rm 2, eq}\left(R_{\rm L}; R_{\rm C}\right)$, which is related to the chemical potential $\mu_{1}\left(R_{\rm L};R_{\rm C}\right)$ of the shell given by Eq.~(\ref{eq:D28}) through
\begin{equation}
\mu_{1}\left(R_{\rm L}; R_{\rm C}\right)=\beta^{-1}\ln\left(c_{\rm 2, eq}\left(R_{\rm L}; R_{\rm C}\right)/c_{2, \infty}\right),
\label{eq:D31}
\end{equation}
which leads to the OF equation given by
\begin{equation}
c_{\rm 2, eq}\left(R_{\rm L}; R_{\rm C}\right) = c_{2, \infty}\exp\left(\beta\mu_{1}\left(R_{\rm L}; R_{\rm C}\right)\right).
\label{eq:D32}
\end{equation}
Note that the concentration $c_{2}\left(R_{\rm L}\right)=c_{\rm 2, eq}\left(R_{\rm L}; R_{\rm C}\right)$ at the boundary $R_{\rm L}$ depends on the radius $R_{\rm C}$.
The equilibrium concentration $c_{\rm 2, eq}\left(R_{\rm L}; R_{\rm C}\right)$ is also expected to decrease monotonically as a function of $R_{\rm L}$ for a fixed $R_{\rm C}$ from Eq.~(\ref{eq:D28}).  At the critical point with $R_{\rm L}^{*}$ and $R_{\rm C}^{*}$, $c_{\rm 2, eq}\left(R_{\rm L}^{*}; R_{\rm C}^{*}\right)=c_{2, \infty}$ because $\mu_{1}\left(R_{\rm L}^{*}; R_{\rm C}^{*}\right)=0$ from Eq.~(\ref{eq:D28}).

When the core nucleus C and the surrounding solution L$_{1}$ are always in chemical equilibrium, $c_{1}\left(R_{\rm C}\right)=c_{\rm 1, eq}\left(R_{\rm C}\right)$.  Similarly, when the surrounding spherical shell of solution L$_{1}$ and the mother solution L$_{2}$ are always in chemical equilibrium, $c_{2}\left(R_{\rm L}\right)=c_{\rm 2, eq}\left(R_{\rm L}; R_{\rm C}\right)$. The concentration $c_{1}\left(R_{\rm L}\right)\neq c_{\rm 1, eq}\left(R_{\rm L}\right)$ is determined from the continuity condition of Eqs.~(\ref{eq:D8}) and (\ref{eq:D9}).  Two equilibrium concentrations $c_{\rm 1, eq}\left(R_{\rm C}\right)$ and $c_{\rm 2, eq}\left(R_{\rm L}; R_{\rm C}\right)$ are determined from the implicit equations  Eqs.~(\ref{eq:D27}), (\ref{eq:D30}), (\ref{eq:D28}), and (\ref{eq:D32}) by replacing  $c_{1}\left(R_{\rm C}\right)$ by $c_{\rm 1, eq}\left(R_{\rm C}\right)$ and $c_{2}\left(R_{\rm L}\right)$ by $c_{\rm 2, eq}\left(R_{\rm L}; R_{\rm C}\right)$

In Fig.~\ref{fig:3}, we show a schematic concentration profile $c_{1}\left(r\right)$ and $c_{2}\left(r\right)$ given by Eqs.~(\ref{eq:D5}) and (\ref{eq:D4}) as well as $c_{\rm 1, eq}\left(R_{\rm C}\right)$ and $c_{\rm 2, eq}\left(R_{\rm L}; R_{\rm C}\right)$ (for a fixed $R_{\rm C}$) at the boundaries for the post-critical core-shell critical nucleus with $R_{\rm C}>R_{\rm C}^{*}$ and $R_{\rm L}>R_{\rm L}^{*}$.

For the post-critical nucleus, the concentrations profiles $c_{1}\left(r\right)$ and $c_{2}\left(r\right)$ given by Eqs.~(\ref{eq:D5}) and (\ref{eq:D4}) are not uniform anymore and the depletion zone increases. Since the two interfaces at $R_{\rm C}$ and $R_{\rm L}$ become flat as $R_{\rm C}\rightarrow \infty$ and $R_{\rm L}\rightarrow \infty$, two concentrations $c_{\rm 1, eq}\left(R_{C}\right)$ and $c_{\rm 2, eq}\left(R_{\rm L}; R_{\rm C}\right)$ at the boundaries must approach their saturation concentration at the flat interface $c_{\rm 1, sat}$ and $c_{\rm 2, sat}$.  Therefore, we have
\begin{eqnarray}
c_{\rm 1, eq}\left(R_{\rm C}\right) \rightarrow c_{\rm 1, sat}
\label{eq:D33} \\
c_{\rm 2, eq}\left(R_{\rm L}; R_{\rm C}\right) \rightarrow c_{\rm 2, sat}
\label{eq:D34}
\end{eqnarray}
as shown in Fig.~\ref{fig:3}

When the radii $R_{\rm C}$ and $R_{\rm L}$ are close to the critical radii $R_{\rm C}^{*}$ and $R_{\rm L}^{*}$, we can replace $c_{2}\left(R_{\rm L}\right)$ by $c_{\rm 2, eq}\left(R_{\rm L}; R_{\rm C}\right)$ in Eqs.~(\ref{eq:D8}) and (\ref{eq:D9}), and expand exponential in Eq.~(\ref{eq:D32}).  Then the driving forces $j_{R_{\rm C}}$ and $j_{R_{\rm L}}$ of growing nucleus radii are given by
\begin{eqnarray}
j_{R_{\rm C}} 
&=& D_{2}\left(\frac{R_{\rm L}}{R_{\rm C}}\right)\frac{\beta c_{2, \infty}\mu_{1}\left(R_{\rm L}; R_{\rm C}\right)}{R_{\rm C}}
\label{eq:D35} \\
j_{R_{\rm L}}
&=& D_{2}\frac{\beta c_{2, \infty}\mu_{1}\left(R_{\rm L}; R_{\rm C}\right)}{R_{\rm L}}
\label{eq:D36} 
\end{eqnarray}
Not only the growth velocity $dR_{\rm C}/dt$ but also $dR_{\rm L}/dt$ vanish at the critical point when $R_{\rm C}=R_{\rm C}^{*}$ and $R_{\rm L}=R_{\rm L}^{*}$ because $\mu_{1}\left(R_{\rm C}^{*}, R_{\rm L}^{*}\right)=0$ from Eq.~(\ref{eq:D28}).

Equations (\ref{eq:D15}) and (\ref{eq:D16}) combined with Eqs. (\ref{eq:D35}) and (\ref{eq:D36}) are similar in form to the Zeldovich relation~\cite{van Putten2009,Iwamatsu2014} derived from the thermodynamic evolution equation~\cite{van Putten2009,Slezov2009,Iwamatsu2014}.  Therefore, the Zeldovich relation can be equivalent to the diffusional growth equation in our core-shell nucleus as well.  However, this is valid only near the thermodynamic critical point with critical radii $R_{\rm C}^{*}$ and $R_{\rm L}^{*}$, as it is recognized that the expansion of the exponential in Eq.~(\ref{eq:D32}) is valid only near the critical radius, and the full expression in Eqs.~(\ref{eq:D30}) and (\ref{eq:D32}) will be necessary to describe the subsequent growth of the nucleus~\cite{Mantzaris2005}.

Now, we can discuss the evolution of this composite core-shell nucleus from Eqs.~(\ref{eq:D35}) and (\ref{eq:D36}).  When the concentration is uniform and oversaturated ($c_{1}\left(r\right)=c_{1, \infty}>c_{\rm 1, sat}$ and $c_{2}\left(r\right)=c_{2, \infty}>c_{\rm 2, sat}$), the core radius $R_{\rm C}$ and the outer shell radius $R_{\rm L}$ reach the critical radius $R_{\rm C}^{*}$ and $R_{\rm L}^{*}$ simultaneously.  Therefore, there is only a {\it single critical point} which corresponds to a {\it single critical nucleus} characterized by {\it two critical radii} $R_{\rm C}^{*}$ and $R_{\rm L}^{*}$.  After passing through this single critical point characterized by {\it a single energy barrier}, two radii $R_{\rm C}$ and $R_{\rm L}$ of the core-shell nucleus start to increase. The growth of the nucleus follows exactly the same scenario given by CNT, by regarding the $R_{\rm C}$ and $R_{\rm L}$ as the radius of nucleus.

In the limit of $R_{\rm C}\rightarrow \infty$ and $R_{\rm L}\rightarrow \infty$, we have
\begin{eqnarray}
j_{R_{\rm C}} &\rightarrow& -D_{2}\frac{c_{2,\infty}-c_{\rm 2, sat}}{R_{\rm C}},
\label{eq:D37} \\
j_{R_{\rm L}} &\rightarrow& -D_{2}\frac{c_{2,\infty}-c_{\rm 2, sat}}{R_{\rm L}},
\label{eq:D38}
\end{eqnarray}
from Eqs.~(\ref{eq:D8}), (\ref{eq:D9}) and (\ref{eq:D34}).  Then the evolution equations Eqs.~(\ref{eq:D15}) and (\ref{eq:D16}) are written as
\begin{eqnarray}
R_{\rm C}\frac{dR_{\rm C}}{dt}=v_{\rm m}D_{2}\left(c_{2,\infty}-c_{\rm 2, sat}\right),
\label{eq:D39} \\
R_{\rm L}\frac{dR_{\rm L}}{dt}=v_{\rm m}D_{2}\left(c_{2,\infty}-c_{\rm 2, sat}\right).
\label{eq:D40}
\end{eqnarray}
Therefore, two radii grow according to the classical law~\cite{Zener1949,Grinin2004} $R_{\rm C}\propto \sqrt{t}$ and  $R_{\rm L}\propto \sqrt{t}$.  Furthermore $R_{\rm L}^{2}-R_{\rm C}^{2}\rightarrow {\rm constant}$.  Then, the radius $R_{\rm C}$ eventually catches up $R_{\rm L}$ as $R_{\rm C}\rightarrow \infty$ and the surrounding shell of solution L$_{1}$ disappears as the Ostwald's step rule predicted~\cite{Ostwald1897,Santra2013}.

\section{\label{sec:sec4}Discussion and Conclusion}

In this paper, we considered the evolution of a core-shell-type nucleus.  The diffusional growth equations both for the core nucleus and the surrounding shell lead to two evolution equations similar to the Zeldovich equation of classical nucleation theory (CNT). Therefore, the kinetic critical radii for both the core radius and the surrounding shell radius can be defined, and there must be a single core-shell critical nucleus which corresponds to a single critical point and a single energy barrier.  When diffusion is so fast that the concentrations at the two surfaces are maintained at the value given by the equilibrium Ostwald-Freundlich boundary condition, simple expressions similar to those for the thermodynamic critical radius of CNT can be derived.  Our formulation further predicts a growing post critical nucleus with a core nucleus surrounded by a metastable intermediate solution, whose thickness will be decreased during the course of evolution in accordance with the Ostwald's step rule.  

Although our discussion takes into account the two-step diffusion of a composite nucleus, a more thorough consideration not only of the diffusion processes but also of the reaction and the reorganization processes within the bulk crystal nucleus is required to understand the growth of near-spherical solid nucleus, since simple diffusional attachment is known to lead to non-spherical fractal structures~\cite{Liu1990,Meakin1993}.  Numerical simulations such as the kinetic Monte Carlo method~\cite{Gorshkov2009,Gorshkov2010} hold the greatest potential for the study of such reorganization processes.  Finally, since we considered only the material diffusion and neglected the heat flow, the instability and fractal growth~\cite{Mullins1963} of the nucleus caused by thermal diffusion cannot be discussed within our present formalism.

\begin{acknowledgments}
This work was partially supported under a project for strategic advancement of research infrastructure for private universities, 2015-2020, operated by MEXT, Japan. 
\end{acknowledgments}

\appendix*

\section{}
Here, we have provided the results for the two-dimensional (2D) circular core-shell nucleus.  Main differences from the three-dimensional (3D) spherical nucleus of the main text are the mathematical forms of the solution of diffusion equation and the thermodynamic Ostwald-Freundlich condition. 

The solution of the 2D diffusion equation Eq.~(\ref{eq:D3}) are written as~\cite{Chakraverty1967}
\begin{eqnarray}
c_{1}\left(r\right) &=&
\frac{c_{1}\left(R_{\rm L}\right)-c_{1}\left(R_{\rm C}\right)}{\ln R_{\rm L}-\ln R_{\rm C}}\ln r \nonumber \\
&&+ \frac{c_{1}\left(R_{\rm C}\right)\ln R_{\rm L}-c_{1}\left(R_{\rm L}\right)\ln R_{\rm C}}{\ln R_{\rm L}-\ln R_{\rm C}}
\nonumber \\
&& R_{\rm C}<r<R_{\rm L}
\label{eq:A1} \\
c_{2}\left(r\right) &=&
\frac{c_{2, \infty}-c_{2}\left(R_{\rm L}\right)}{\ln R_{\infty}-\ln R_{\rm L}}\ln r \nonumber \\
&&+ \frac{c_{2}\left(R_{\rm L}\right)\ln R_{\infty}-c_{2, \infty}\ln R_{\rm L}}{\ln R_{\infty}-\ln R_{\rm L}}
\nonumber \\
&& R_{\rm L}<r<R_{\infty}
\label{eq:A2} 
\end{eqnarray}
where $R_{\infty}$ is the large radius from the nucleus and $c_{2, \infty}$ is the concentration at this distance.

The free energy of the core-shell nucleus in the 2D case that corresponds to Eq.~(\ref{eq:D18}) is written as
\begin{equation}
G\left(n_{\rm c},n_{1}\right)
=-n_{\rm C}\Delta\mu_{1}-n_{1}\Delta\mu_{2}
+2\pi R_{\rm C}\gamma_{c,1}+2\pi R_{\rm L}\gamma_{1,2},
\label{eq:A3}
\end{equation}
with
\begin{equation}
n_{\rm C}=\frac{\pi}{v_{\rm m}}R_{\rm C}^{2}
\label{eq:A4}
\end{equation}
and
\begin{equation}
n_{1}=\int_{R_{\rm C}}^{R_{\rm L}}2\pi r c_{1}\left(r\right) dr
\label{eq:A5}
\end{equation}
Following the same procedure as that of Section III, the critical radius $R_{\rm C}^{*}$ which corresponds to Eq.~(\ref{eq:D23}) is given by
\begin{equation}
R_{\rm C}^{*}=\frac{\gamma_{\rm c,1}v_{\rm m}}{\Delta\mu_{\rm 1,eff}},
\label{eq:A6}
\end{equation}
where $\Delta\mu_{\rm 1, eff}$ is given by Eq.~(\ref{eq:D24}).  The critical radius $R_{\rm L}^{*}$ of the intermediate metastable solution is give by a formula
\begin{equation}
R_{\rm L}^{*}=\frac{\gamma_{1,2}}{c_{1, \infty}\Delta\mu_{2}}.
\label{eq:A7}
\end{equation}
similar to Eq.~(\ref{eq:D26}).  Therefore, Eqs.~(\ref{eq:A6}) and (\ref{eq:A7}) can be derived from Eqs.~(\ref{eq:D23}) and (\ref{eq:D26}) by replacing $2\gamma_{c,1}$ by $\gamma_{c, 1}$ and $2\gamma_{1,2}$ by $\gamma_{1,2}$.  Using the same replacement in Eqs.~(\ref{eq:D27}) and (\ref{eq:D28}), we can derive the formula for the chemical potentials $\mu_{\rm C}\left(R_{\rm C}\right)$ and $\mu_{\rm L}\left(R_{\rm C}, R_{\rm L}\right)$ in two dimensional case.  The evolution of the nucleus is described by Eqs.~(\ref{eq:D15}) and (\ref{eq:D16}) with Eqs.~(\ref{eq:D35}) and (\ref{eq:D36}).  Therefore, the scenario of nucleation and growth of the two-dimensional core-shell nucleus is the same as that of the three-dimensional nucleus.


\begin{thebibliography}{99}
\bibitem{Kelton2010} K. F. Kelton and A. L. Greer, Nucleation in Condensed Matter, Applications in Materials and Biology, Pergamon, Oxford, 2010, Chapter 6.
\bibitem{Gorshkov2010} V. Gorshkov and V. Privman, Physica E {\bf 43}, 1 (2010).
\bibitem{Meldrum2008} F. C. Meldrum and H. C\"ofen, Chem. Rev. {\bf 108}, 4332 (2008).
\bibitem{Sear2012} R. P. Sear, Int. Mater. Rev. {\bf 57}, 328 (2012).
\bibitem{Vekilov2012} P. G. Vekilov, J. Phys.: Condens Matter {\bf 24}, 193101 (2012).
\bibitem{Zener1949} C. Zener, J. Appl. Phys. {\bf 20}, 950 (1949).
\bibitem{Frank1950} F. C. Frank, Proc. R. Soc. London Ser. A {\bf 201}, 586 (1950).
\bibitem{Epstein1950} P. S. Epstein and M. S. Plesset, J. Chem. Phys. {\bf 18}, 1505 (1950).
\bibitem{Reiss1951} H. Reiss, J. Chem. Phys. {\bf 19}, 482 (1951).
\bibitem{Slezov2009} V. V. Slezov, Kinetics of First-Order Phase Transition, Wiley-VCH, Weinheim, 2009, Chapter 5.
\bibitem{Wen2014} T. Wen, L. N. Brush, and K. M. Krishnan, J. Colloid. Interface Sci. {\bf 419}, 79 (2014).
\bibitem{Robb2008} D. T. Robb and V. Privman, Langmuir {\bf 24}, 26 (2008).
\bibitem{Grinin2011} A. P. Grinin, G. Yu. Gor, and F. M. Kuni, Atoms. Res. {\bf 101}, 503 (2011).
\bibitem{Lutsko2011} J. F. Lutsko, J. Chem. Phys. {\bf 135}, 161101 (2011).
\bibitem{Peters2011} B. Peters, J. Chem. Phys. {\bf 135}, 044107 (2011).
\bibitem{Iwamatsu2014} M. Iwamatsu, J. Chem. Phys. {\bf 140}, 064702 (2014).
\bibitem{Zeldovich1943} Ya. B. Zeldovich, Acta Physicochim URSS {\bf 18}, 1 (1943).
\bibitem{van Putten2009} D. S. van Putten and V. I. Kalikmanov, J. Chem. Phys. {\bf 130}, 164508 (2009).
\bibitem{tenWolde1997} P. R. ten Wolde and D. Frenkel, Science {\bf 277}, 1975 (1997).
\bibitem{Heijna2007} M. C. R. Heijna, W. J. P. van Enckevort, and E. Vlieg, Phys. Rev. E {\bf 76}, 011604 (2007).
\bibitem{Sleutela2014} M. Sleutela, and A. E. S. Van Driesscheb, Proc. Natl. Acad. Sci. {\bf 111}, E546 (2014).
\bibitem{Iwamatsu2011} M. Iwamatsu, J. Chem. Phys. {\bf 134}, 164508 (2011).
\bibitem{Iwamatsu2012} M. Iwamatsu, Phys. Rev. E {\bf 86}, 041604 (2012).
\bibitem{Wang2016} W. L. Wang, Y. H. Wu, L. H. Li, D. L. Geng, and B. Wei, Phys. Rev. E {\bf 93}, 032603 (2016).
\bibitem{Desgranges2007} C. Desgranges and J. Delhommelle, J. Am. Chem. Soc. {\bf 129}, 7012 (2007).
\bibitem{Meel2008} J. A. van Meel, A. J. Page, R. P. Sear, and D. Frenkel, J. Chem. Phys. {\bf 129}, 204505 (2008).
\bibitem{Iwamatsu2010} M. Iwamatsu, J. Alloy. Comp. {\bf 504S}, S538 (2010).
\bibitem{Qi2015} W. Qi, Y. Peng, Y. Han, R. K. Bowles, and M. Dijkstra, Phys. Rev. Lett. {\bf 115}, 185701 (2015).
\bibitem{Kelton2000} K. F. Kelton, Acta Mater. {\bf 48}, 1967 (2000).
\bibitem{Granasy2000} L. Granasy and D. W. Oxtoby, J. Chem. Phys. {\bf 112}, 2410 (2000).
\bibitem{Santra2013} M. Santra, R. S. Singh, and B. Bagchi, J. Phys. Chem. B {\bf 117}, 13154 (2013).
\bibitem{Lutsko2016} J. F. Lutsko, J. Phys.: Condens. Matter {\bf 28}, 244020 (2016).
\bibitem{Ostwald1897} W. Ostwald, Z. Phys. Chem. {\bf 22}, 289 (1897).
\bibitem{Grinin2004} A. P. Grinin, A. K. Shchekin, F. M. Kuni, E. A. Grinina, and H. Reiss, J. Chem. Phys. {\bf 121}, 387 (2004).
\bibitem{Mantzaris2005} N. V. Mantzaris, Chem. Eng. Sci. {\bf 60}, 4749 (2005).
\bibitem{Meakin1993} P. Meakin, Phys. Rep. {\bf 235}, 189 (1993).
\bibitem{Liu1990} F. Liu and N. Goldenfeld, Phys. Rev. A {\bf 42}, 895 (1990).
\bibitem{Gorshkov2009} V. Gorshkov, A. Zavalov, and V. Privman, Langmuir {\bf 25}, 7940 (2009).
\bibitem{Mullins1963} W. W. Mullins and R. F. Sekerka, J. Appl. Phys. {\bf 34}, 323 (1963). 
\bibitem{Chakraverty1967} B. K. Chakraverty, J. Phys. Chem. Sol. {\bf 28}, 2401 (1967).
\end{thebibliography}

\end{document}